\documentclass{article}

\usepackage{arxiv}

\usepackage[utf8]{inputenc} 
\usepackage[T1]{fontenc}    
\usepackage{hyperref}       
\usepackage{url}            
\usepackage{booktabs}       
\usepackage{amsfonts}       
\usepackage{nicefrac}       
\usepackage{microtype}      
\usepackage{lipsum}		
\usepackage{graphicx}
\usepackage{natbib}
\usepackage{doi}

\title{Whisper in Focus: Enhancing Stuttered Speech Classification with Encoder Layer Optimization}

\author{ 
{Huma Ameer} \\
	School of Electrical Engineering and\\
 Computer Science (SEECS)\\
	National University of Sciences\\
 and Technology (NUST)\\
	Islamabad, Pakistan \\
	\texttt{hameer.msds20seecs@seecs.edu.pk} \\
	\And
  {Seemab Latif} \\
	School of Electrical Engineering and\\
 Computer Science (SEECS)\\
	National University of Sciences\\
 and Technology (NUST)\\
	Islamabad, Pakistan \\
	\texttt{seemab.latif@seecs.edu.pk} \\
 \And
 {Rabia Latif} \\
	College of Computer and \\
 Information Sciences (CCIS)\\
    Prince Sultan University\\	
	Riyadh, Saudi Arabia \\
	\texttt{rlatif@psu.edu.sa} \\
\And
	{Sana Mukhtar} \\
	Shaheed Zulfikar Ali Bhutto Institute\\
 of Science and Technology (SZABIST)\\
	Islamabad, Pakistan \\
	\texttt{sana@szabist-isb.edu.pk} \\
}

\date{}

\renewcommand{\headeright}{}
\renewcommand{\undertitle}{}
\renewcommand{\shorttitle}{Enhancing Stuttered Speech Classification with Encoder Layer Optimization}

\hypersetup{
pdftitle={Enhancing Stuttered Speech Classification with Encoder Layer Optimization},
pdfsubject={q-bio.NC, q-bio.QM},
pdfauthor={David S.~Hippocampus, Elias D.~Striatum},
pdfkeywords={First keyword, Second keyword, More},
}

\begin{document}
\maketitle

\begin{abstract}
	In recent years, advancements in the field of speech processing have led to cutting-edge deep learning algorithms with immense potential for real-world applications. The automated identification of stuttered speech is one of such applications that the researchers are addressing by employing deep learning techniques. Recently, researchers have utilized Wav2vec2.0, a speech recognition model to classify disfluency types in stuttered speech. Although Wav2vec2.0 has shown commendable results, its ability to generalize across all disfluency types is limited. In addition, since its base model uses 12 encoder layers, it is considered a resource-intensive model. Our study unravels the capabilities of Whisper for the classification of disfluency types in stuttered speech. We have made notable contributions in three pivotal areas: enhancing the quality of SEP28-k benchmark dataset, exploration of Whisper for classification, and introducing an efficient encoder layer freezing strategy. The optimized Whisper model has achieved the average F1-score of 0.81, which proffers its abilities. This study also unwinds the significance of deeper encoder layers in the identification of disfluency types, as the results demonstrate their greater contribution compared to initial layers. This research represents substantial contributions, shifting the emphasis towards an efficient solution, thereby thriving towards prospective innovation.
\end{abstract}


\label{sec:introduction}
Speech disfluency is a disruption in the normal flow of speech, a universal phenomenon encountered by most speakers. It is quite common to repeat words or use filler words during conversation. Stuttering, however, is a speech disorder associated with various disfluency types. There are certain evaluation criteria and factors which are identified by Speech Language Pathologists (SLP) while assessing stuttering in their patients. Stuttering affects almost 1\% of the population worldwide \cite{yairi2013epidemiology}. Notably, the common disfluency types that stuttering manifests through are; blocks, prolongation, interjection, and sound/word repetitions \cite{shipley2019assessment}. 
Table~\ref{tab:stutter-types} provides an explanation for these labels.

\begin{table}[htbp]
\centering
\caption{Common Disfluency types}
\label{tab:stutter-types} 
\begin{tabular}{l|l}
\toprule
\textbf{Disfluency Type} & \textbf{Example} \\
\midrule
Blocks & I was \underline{\textbf{blockage/pause}} sleeping \\
Prolongations & I was \underline{\textbf{ssss}}sleeping \\
Interjection & I \underline{\textbf{uh}} was \underline{\textbf{uhm uh}} sleeping \\
Sound Repetition & I \underline{\textbf{w-w-w-w}}as sleeping \\
Word Repetition & I \underline{\textbf{was was}} sleeping \\
\bottomrule
\end{tabular}
\end{table}

The evaluation of stuttered speech necessitates the identification of the aforementioned disfluency types. In conventional evaluation methods, SLPs have to vigilantly analyze the speech samples, and calculate the parameters such as the total number of disfluencies, frequencies of different types of disfluencies, speech rate etc \cite{shipley2019assessment}. Therefore, the procedure for assessment is time-intensive and it invloves the risk of biases-decision making. Given the complexity of the assessment process, we aim to propel the field forward through the automation of disfluency type identification. Our ultimate goal is to create a system that not only streamlines but also significantly enhances the efficiency of therapists’ assessment procedures.

The automated stuttering identification can be leveraged through various data modalities,including bio-respiratory signals \cite{villegas2019novel}, visual data \cite{yildirim2009automatic}, functional near-infrared spectroscopy (fNIRS) \cite{hosseini2018fnirs,chang2014research}, textual data \cite{geetha2000classification}, and speech data \cite{sheikh2021stutternet}. Among these modalities, speech data emerges as the most practical and cost-effective means of data collection.  

In the realm of automated stutter detection, there have been significant advancements evolving from acoustic features extractions paired with machine learning techniques \cite{mahesha2013classification,tan2007application,chee2009mfcc} to the adoption of deep learning approaches \cite{santoso2019classification,kourkounakis2020detecting}. Recently, the adoption of transformer-based models like Wav2vec2.0 \cite{baevski2020wav2vec} have demonstrated encouraging results \cite{sebastian2022detecting,bayerl2023classification}. Nevertheless, the quest for models that are both generalized and resource-efficient persists as a challenge at hand. In our study, we have delved into the applicability of a weakly supervised model known as Whisper \cite{radford2023robust} developed by Open AI \footnote{Open AI Whisper: \url{https://openai.com/research/whisper}}. It has exhibited remarkable performance in contrast to Wav2vec2.0, across various tasks, including speech recognition, audio classification, and transcription. In addition we have designed and tested different fine-tuning strategies on a well-establised dataset known as Stuttering Events in Podcasts (SEP-28k) \cite{lea2021sep}. These strategies were tested on an FluencyBank \cite{ratner2018fluency}.

This study's contributions are threefold:

\begin{itemize}
    \item We have introduced a refined version of the SEP-28k dataset, with enhanced quality and utility as a valuable resource in the field.
    \item We have explored the applicability of Whisper, which, to the best of our knowledge, marks the first investigation of its potential in the classification of disfluency types. Our findings demonstrate that Whisper yields improved results, which is a significant advancement in this domain.
    \item The investigation of freezing and unfreezing of encoder layers has unveiled a noteworthy strategy that reduces the number of learnable parameters and significantly enhances efficiency. Thus, this firmly establishes Whisper as an optimized choice for this task.
\end{itemize}

This paper is divided into 5 sections: Section~\ref{sec:introduction} briefly discusses the essential background and context of our research. Section ~\ref{sec:literature-review} presents a comprehensive literature review, which highlights the research studies that have contributed significantly to the classification of disfluency types. Moving on to the methodology section ~\ref{sec:methodology},  we have explained data refinement procedures, and the overall experimentation process employed in our study. The results and discussion section ~\ref{sec:results-discussion}, not only showcases the findings of our research but also provide an insightful analysis and interpretation of these outcomes. Finally, section ~\ref{sec:conclusion-future} serves as the conclusion and future directions, which summarizes our key findings and discusses future research endeavors.

\section{Literature Review}
\label{sec:literature-review}
The problem of stuttered speech classification has been approached through various methods ranging from statistical machine learning methods to deep learning methods \cite{sheikh2022machine}. The first step in classification of stuttered speech is feature extraction of the audio clips. The  conventional feature extracting techniques i.e. Mel Frequency Cepstral Coefficients (MFCC), Linear Predictive Coding (LPC) are used, however these techniques donot provide learnable features. This necessitates the usage of learnable feature extractors i.e. Wav2vec2.0, Whisper etc. 

For classification of disfluencies in stuttered speech techniques such as Hidden Markov Model (HMM) have been employed \cite{noth2000automatic,sheikh2022machine}. Since these methods have simple architectures, their capabilities are limited when dealing with non linear data. Therefore, it is difficult to classify disfluency types in stuttered speech as this unstructured data is complex in nature \cite{sheikh2022machine}. In addition, these  approaches require manual feature engineering which makes it cumbersome to deal with the particular tasks. Deep learning techniques automates this process with improved outcomes.  In this section we will uncover the superior results achieved using Wav2vec2.0 on single disfluencies. Next, we will discuss the architecture and potential applications of Whisper. Lastly, the datasets in focus will be discussed.
 \subsection{Wav2vec2.0}
\begin{figure}[h]
    \centering
     \includegraphics[width=4.5in]{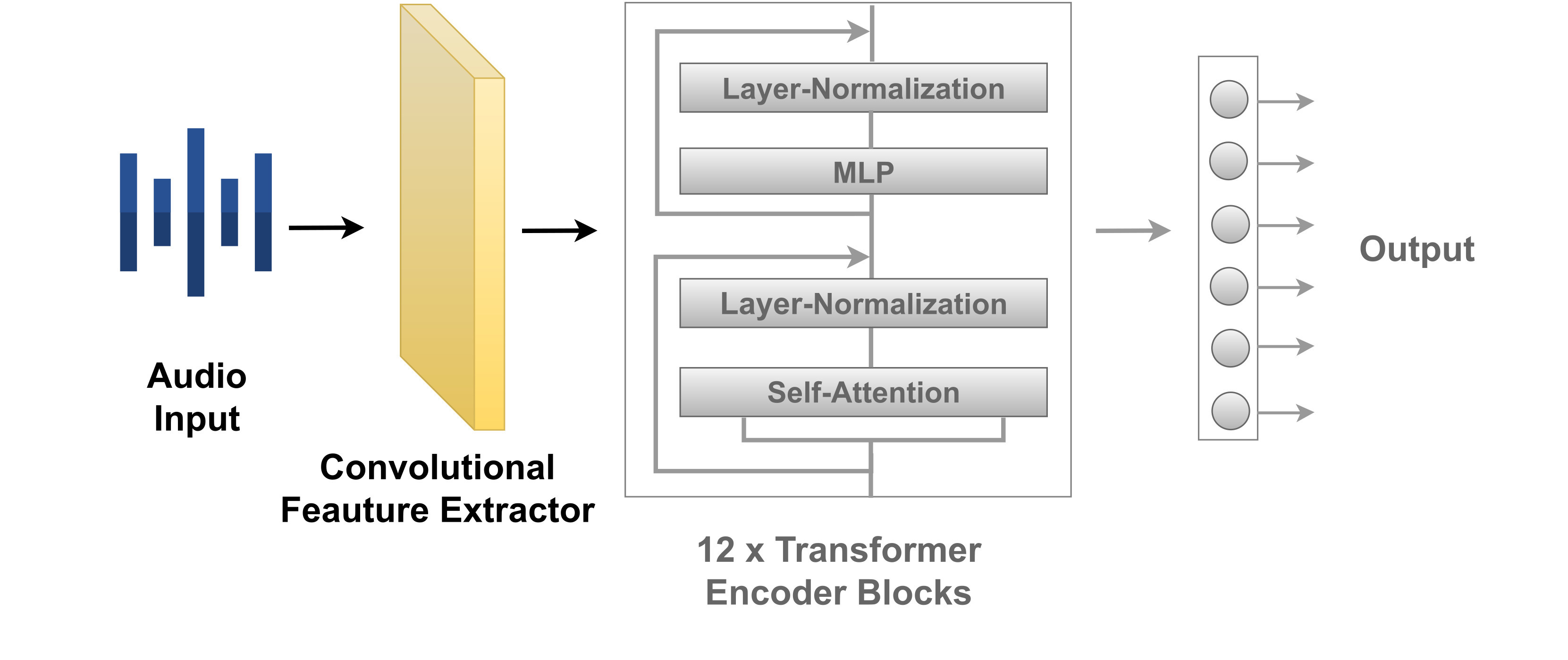}
    \caption{Wav2vec2.0 Encoder Overview}
    \label{fig:wav2vec2}
\end{figure}

Recently, transformer-based architecture i.e. Wav2vec2.0 has been shown to achieve groundbreaking results for the classification of disfluencies considering the complexity of the problem \cite{baevski2020wav2vec}. Wav2vec2.0 is a self-supervised learning framework, which encodes speech from convolutional neural networks to learn the representations of raw audio as shown in Figure~\ref{fig:wav2vec2}. These representations are fed to the encoder of a transformer model to extract contextual embeddings. The decoder is then used for the generation of text depending on the task at hand. Initially, the model is trained on unlabeled speech data, through fine-tuning with labelled data, it is adaptable for various downstream applications.

The authors \cite{bayerl2022influence}, have demonstrated the influence of splits of the datasets SEP-28k, and reported the results on various splits. Wav2vec2.0 has been used for feature extraction with the classifier Support Vector Machine (SVM) for the detection of disfluencies in speech. The highest F1-score achieved for each type of dysfluency varies as it depends on the model and splits used. For example, the highest F1-score for \textit{Block} disfluency was 0.48 using SVM, whilst, for Interjection, the highest score achieved was 0.71. The notion of experimenting with various splits is commendable, as this data-centric approach can be utilized to illustrate the capabilities of the model. However, the study has not presented any work on the quality of the dataset.

Using Wav2vec2.0 for feature extraction, the authors have proposed a Convolutional Neural Network (CNN) for identifying the disfluencies in speech \cite{mohapatra2022speech}. The authors have illustrated the importance of data quality over quantity in their research study. The utilization of high-quality data with high inter-annotator agreement, balanced classes, and contextual embeddings from a pretrained network is beneficial for disfluency classification models. Their proposed architecture, DisfluencyNet incorporates these techniques to improve the performance in automated stuttered speech detection. The model achieves high performance with only a quarter of the data for training, which demonstrates its efficiency and effectiveness. The model is evaluated with various sizes of training data, and the performance is compared to baseline results from SEP-28k and FluencyBank datasets. The proposed methodology achieved an F1-score greater than 0.7 for all types of disfluencies except blocks. The suggested technique is applied on individual disfluencies only. Another study has taken the same data distribution approach proposed by \cite{liu2023automatic}. The authors have used speech embeddings from Wav2vec2.0, followed by transformer and CNNs. This model architecture was applied on the SEP-28k dataset, and a Chinese dataset referred to as PSC-PS-DF. The technique considers the issue of disfluent speech with variable-length, modified by entropy invariance of attention mechanism. But this model is utilized for single task detection which is not applicable in the real world.

Similarly, \cite{bayerl2022ksof} have put forth a dataset in German language, keeping the SEP-28k as a base dataset. The proposed dataset included annotations for 6 disfluency types, including blocks, prolongations, sound repetitions, word repetitions, modified, and interjections. For the classifications of these types, authors have experimented with 3 feature extraction techniques, Opensmile, Wav2vec2.0, and MFCC. For classification, SVM is utilized, and for MFCC as a feature extractor, authors have experimented with LSTM, and LSTM with Attention. The results were evaluated on the F1-score. It was deduced that Wav2vec2.0 performed the best because of its capability to extract the intricacies of speech. The best F1-score of classes, modified, blocks, interjections, prolongation, sound repetitions, and word repetitions were 0.73, 0.57, 0.59, 0.40, 0.43, and 0.17, respectively. However, the authors took this problem as a binary classification and experimented on each disfluency type separately. This implies that the proposed models may not necessarily be one for all.

The work presented by Bayeral et al. has demonstrated the effectiveness of fine-tuning Wav2vec2.0 for the classification of disfluencies in stuttered speech \cite{bayerl22_interspeech}. The concept of multi-task learning was used in this study for the purpose of regularization. The proposed method with Wav2vec2.0 as a feature extractor and SVM for classification was evaluated on KSoF and Fluency Bank dataset. The highest F1-score achieved was 0.84 for \textit{Interjection} on Fluency Bank and 0.76 for \textit{Modified Speech} on KSoF. The results show that using embeddings from the fine-tuned models can lead to significant improvements in classification performance, with relative gains of up to 27\% in F1-score.
\subsection{Whisper}

\begin{figure}[h]
    \centering
    \includegraphics[width=\linewidth]{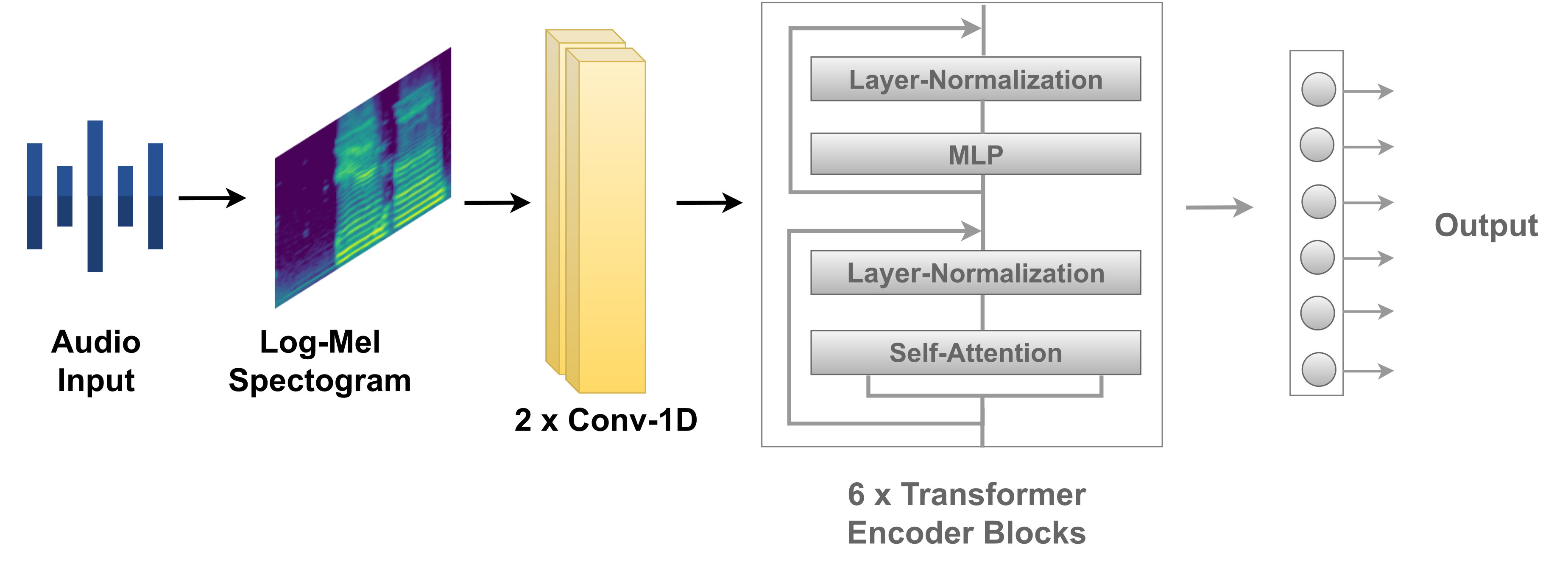}
  
    \caption{Encoder-Centric Whisper Architecture}
    \label{fig:whisper}
\end{figure}

 In contrast to Wav2vec2.0, Whisper is a weakly supervised transformer-based encoder-decoder architecture that stands as a current state-of-the-art, specifically for speech recognition \cite{radford2023robust}. Whisper takes the input audio clips with a sampling rate of  16 kHz. These audio clips are then converted to log-Mel spectrogram representation as shown in Figure~\ref{fig:whisper}. Then they are subsequently processed through two 1-D convolutional layers to extract features. These features are then fed into an encoder, and decoder depending on the nature of specific tasks. The decoder is trained on multitask format so that the model can be leveraged for various use cases i.e. speech recognition, language identification, voice detection, and speech translation. 

 The Whisper model is being utilized in various research studies. For instance, Yang et al. \cite{yang2023chinese} harnessed this model to enhance Chinese automatic speech recognition and name entity recognition. In another study, Whisper is leveraged as an audio-encoder for the detection of DeepFake surpassing other speech embedding methods \cite{kawa2023improved}. In research work by Kodali et al. \cite{kodali2023classification}, the encoder of Whisper has been investigated for the automated classification of vocal intensity. The outcomes of Whisper demonstrated superior performance compared to Wav2vec2.0. Similarly, the comparison of Wav2vec2.0 and Whisper for the classification of nonverbal vocalization illustrated the potential of the Whisper model solidifying its capabilities. \cite{tzirakis2023large}. 
\subsection{Datasets}
\label{subsec:datasets}
In the existing literature, two primary datasets SEP-28k \cite{lea2021sep} and FluencyBank \cite{ratner2018fluency} have been leveraged for training and testing the models. SEP-28k consists of ~28k instances, and each audio clip is 3 seconds long. These clips were extracted from eight podcasts and labeled with five major disfluency types; prolongation, sound repetition, word repetition, interjections, and blocks. Similarly, FluencyBank has been labelled and consists of the same audio length as SEP-28k. It consists of ~4k instances and is mostly leveraged for testing purposes. Lastly, both datasets consist of audio of English speakers.

As discussed in the previous study presented by \cite{bayerl2022influence}, three speaker-exclusive dataset splits have been proposed i.e. SEP-28k-E, SEP-28k-E-T and SEP-28k-D. The training split of SEP-28k-E consists of clips with the top four dominant speakers. The remaining clips are partitioned for testing and validation. SEP-28k-T and SEP-28k-D are similar as they are using the testing and validation split of SEP-28k-E in their training splits, and testing them on the top four dominant speakers. The splits of SEP-28k-E can exhibit the capability of a model by training them on a few speakers with more audio clips and testing on various speakers. On the other hand, SEP-28k-T and SEP-28k-D can identify the ability of the model to learn from fewer examples of various speakers.
\subsection{Literature Analysis}
By reviewing existing literature, it is evident that transformer-based models have consistently demonstrated high performance when compared to alternative methods. Notably, the outcomes of Wav2vec2.0 have been exceptional throughout various experiments as mentioned in the aforementioned studies. The model parameters and their impact have not been a subject of focus within these investigations. Also, most of the existing literature has proposed a methodology based on discrete disfluencies. This underscores the necessity for a resilient model that has the ability to classify multiple disfluency classes and is resource-efficient. Moreover, in the aforementioned use cases of Whisper, it can be inferred that it has outperformed Wav2vec2.0. Therefore, this study will leverage Whisper for the classification of disfluency types in stuttered speech.

\section{Methodology}
\label{sec:methodology}
The existing methodologies, as reviewed in the literature, have not emphasized on the resource-efficient parameterization. Additionally, we have identified the need for a  classification model which is capable of handling multiple disfluency classes. To this end, the performance of a relatively new transformer-based model Whisper has also been studied. Whisper has demonstrated superior results to Wav2vec2.0 across various scenarios, as discussed in the literature review. On that account, in this study, the encoder component of the Whisper will be utilized for the classification of disfluency types in stuttered speech. In this paper, SEP-28k and FluencyBank datasets will be used for fine-tuning and testing of Whisper, respectively. We will also assess the impact of freezing and unfreezing layers, a strategy that will lead to a reduced number of trainable parameters compared to the prior methodologies.
\begin{figure}[h]
    \centering
    \includegraphics[width=\linewidth]{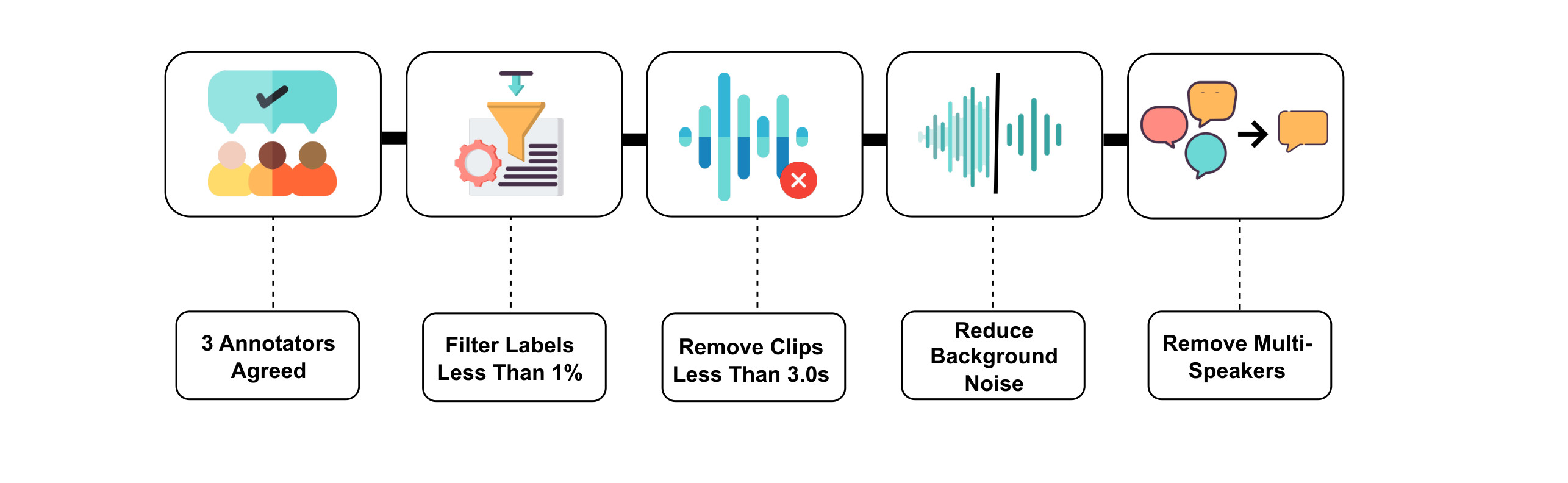}
    \caption{Data Cleaning and Preprocessing Pipeline}
    \label{fig:datacleaning}
\end{figure}
\subsection{Data Cleaning and Preprocessing}
  The study employs the SEP-28k dataset for the classification of disfluency types in stuttered speech. The dataset is a well-established collection for identifying the stuttering events, it consists of  disfluency types i.e. blocks, prolongation, sound repetitions, word repetitions, and interjections. For fine-tuning the transformer-based model, SEP-28k will be used and has been discussed in section \ref{subsec:datasets}.
  To ensure the quality of the dataset, a series of data curation steps were carried out as given in Figure~\ref{fig:datacleaning}. Initially, only the instances in which all three annotators agreed on a specific label were included in the experimentation. The mentioned approach finds its inspiration in the work of Mohapatra et al. \cite{mohapatra2022speech} and has demonstrated the significance of data quality over mere quantity.
  Since the scarcity of labels can hinder the performance of the model as they do not provide sufficient information to learn meaningful patterns. Therefore, in the next step, labels with less than 1\% of representation in the dataset were excluded. These were non-disfluent labels such as Natural Pause, Difficult to Understand, No Speech, Poor Audio Quality, and Music. Furthermore, to uphold consistency, audio clips with durations less than 3 seconds were also excluded. This standardization intends to establish  a baseline duration for all input samples, facilitating more effective model training and evaluation.
  In order to mitigate the potential interference in classification outcome, background noises were removed from the audio clips. The noise removal task was accomplished using a python library noisereduce. Lastly, audio clips consisting of multiple speakers were also omitted as they could introduce complexities in understanding and classification. The audios with background noise and multiple speakers were identified by listening to the clips.
  
\subsection{Data Description}
The three datasets splits proposed by \cite{bayerl2022influence} were fine-tuned to assess the capabilities of the proposed model. The details of these splits are discussed in \ref{subsec:datasets}. In this paper, we have explored an alternative configuration, involving the merger of training and validation of data, with the test split repurposed for validation. These dataset will be referred as SEP-28k-E-merged and SEP-28k-T-merged, their splits are presented in Table \ref{tab:data-split-abbrev}. The model’s generalization capability was assessed by testing it on a distinct FluencyBank dataset. Whisper was applied on the five dataset splits i.e. SEP-28k-E, SEP-28k-T, SEP-28k-D, SEP-28k-E-merged, and SEP-28k-T-merged. 

We have experimented on two different versions of dataset. In the first version we have standardized the dataset by following  the first three preprocessing steps demonstrated in Figure~\ref{fig:datacleaning}. We refer to this version as semi-clean and it has 12,585 instances. Next, in the second version, we have included two more data cleaning steps illustrated in Figure~\ref{fig:datacleaning} such as removing the background noise and removing multi-speakers. 
After including the above mentioned steps, the second version which is referred as clean dataset comprised a total of 12,139 instances. 

\begin{table}[htbp]
\centering
\caption{Data Split Configurations for Experimental Study}
\label{tab:data-split-abbrev}
\renewcommand{\arraystretch}{1.2} 
\begin{tabular}{@{}lll l@{}}
\toprule
Data Split & Training Data & Validation Data & Testing Data \\
\hline
SEP-28k-E & 4-DS & DS-Set 1 & DS-Set 2 \\
SEP-28k-T & DS-Set 1 & DS-Set 2 & 4-DS \\
SEP-28k-D & DS-Set 2 & DS-Set 1 & 4-DS \\
SEP-28k-E-merged & 4-DS + DS-Set 1 & DS-Set 2 & FB \\
SEP-28k-T-merged & DS-Set 1 + DS-Set 2 & 4-DS & FB \\
\bottomrule
\end{tabular}
\vspace{0.01in}
\caption*{Note -- DS = Dominant speakers, DS-Set = Distinct speakers-Set, FB = FluencyBank}
\end{table}
\subsection{Experimentation}
\begin{figure}[h]
    \centering
    \includegraphics[width=\linewidth]{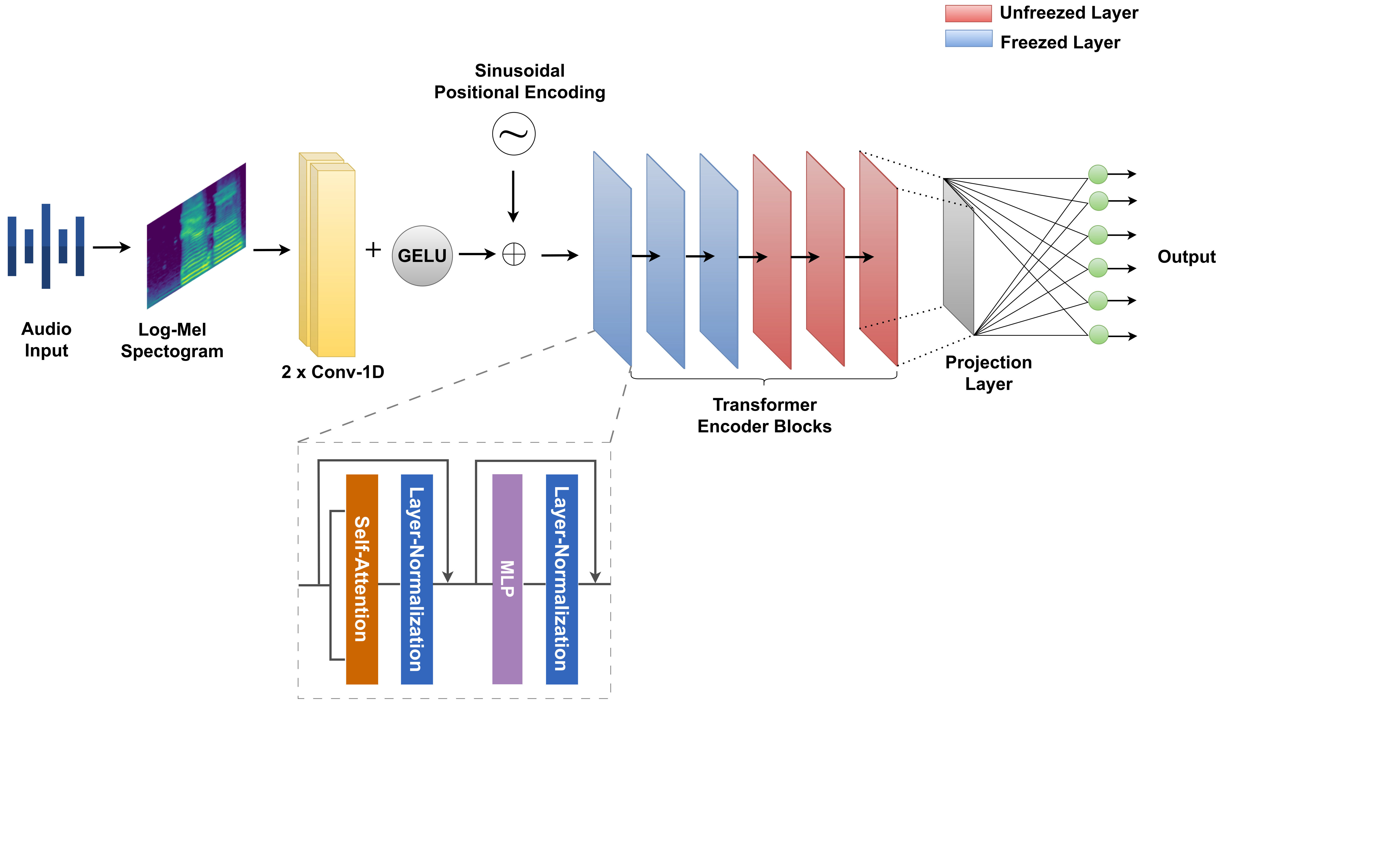}
    \caption{Structural Representation of Whisper Base Model with Proposed Strategy}
    \label{fig:architecture-figure}
\end{figure}

The study capitalizes on the Whisper model to accomplish the classification task. In adapting to this specific classification task, only the encoder component is utilized, followed by a linear layer dedicated for classification as given in Figure~\ref{fig:architecture-figure}. 
The pre-trained English base model is fine-tuned for the targeted downstream task of stuttered speech classification.  For fine-tuning, Whisper for Audio Classification the hugging face library is used\footnote{Hugging Face: \url{https://huggingface.co/docs/transformers/model_doc/Whisper}}.

The F1-score was chosen as our evaluation metric based on the results of prior research. This metric takes into account both precision and recall, which provides a fair evaluation of how well our model performs in classifying disfluency types. The aforementioned five data splits are fine-tuned, on both semi-clean and clean version of data. The splits giving the best F1-score is then chosen for further experimentation. 
The outcomes of SEP-28k-E-merged were substantial, and were comparable in case of both clean and semi-clean data. Therefore, we choose them for further experimentation, where we freezed some encoder layers. The rationale behind this exploration is to utilize the resources more efficiently, as it will reduce the number of trainable parameters while model training. 

Uptill now, we have narrowed down to SEP-28k-E-merged data split with both semi-clean and clean version. The base strategy used here represents the configuration where all the layers are unfreezed. The results obtained using base strategy will be compared with the variation of freezing/unfreezing strategies to further optimize the model. In strategy 1, we are freezing first three layers of the encoders, keeping the last 3 layers unfreezed. Next, in strategy 2, the first 3 layers are unfreezed while freezing the remaining layers .

The models were fine-tuned on the NVIDIA GeForce RTX 2070 GPU, with a batch size of 32, cross entropy as loss function, and the initial learning rate of 2.5 x $10^{-5}$. This specific learning rate is recommended by one of the authors’ of the Whisper paper\footnote{GitHub repository: \url{https://github.com/vasistalodagala/whisper-finetune}}. Lastly, early stopping was also used to avoid model overfitting.

\section{Results and Discussion}
\label{sec:results-discussion}
In this study, our research showcases the effectiveness of a transformer-based model, Whisper for the classification of the disfluency types in stuttered speech as well as strategy for efficient resource utilization. The model is fine-tuned on the SEP-28k-E-merged dataset and then tested on the FluencyBank dataset. The  F1-score of each disfluency type and weighted average F1-score in the Table~\ref{tab:my-result-table} shows the commendable performance of Whisper. 

\begin{table}[htbp]
\centering
\caption{F1-score for Various Strategies on SEP28k-E-Merged Dataset}
\label{tab:my-result-table}
\renewcommand{\arraystretch}{1.2} 
\begin{tabular}{@{}lcccccc@{}}
\toprule
& \multicolumn{3}{c}{\textbf{Semi-cleaned SEP28k-E-Merged}} & \multicolumn{3}{c}{\textbf{Cleaned SEP28k-E-Merged}} \\
\cmidrule(lr){2-4} \cmidrule(lr){5-7}
& Base  & Strategy 1 & Strategy 2 & Base   & \textbf{Strategy 1}   &  Strategy 2  \\
\midrule
\textbf{No Stuttered Words} & 0.27          & 0.18       & 0.16       & 0.16           & \textbf{0.23}         & 0           \\
\textbf{Word Repetition}          & 0.72          & 0.7        & 0.67       & 0.73           & \textbf{0.72}         & 0.68        \\
\textbf{Sound Repetition}         & 0.89          & 0.9        & 0.84       & 0.88           & \textbf{0.89}         & 0.84        \\
\textbf{Prolongation}     & 0.73          & 0.68       & 0.59       & 0.71           & \textbf{0.73}         & 0.63        \\
\textbf{Interjection}     & 0.72          & 0.71       & 0.5        & 0.68           & \textbf{0.7}          & 0.51        \\
\textbf{Block}            & 0.67          & 0.69       & 0.41       & 0.74           & \textbf{0.72}         & 0.35       \\
\midrule
\textbf{Average F1-score} & 0.8 & 0.8 & 0.71 & 0.8 & \textbf{0.81} & 0.71 \\
\bottomrule
\end{tabular}
\end{table}

The experiment yielded promising results, demonstrating the impact of layer freezing strategies for this classification task. Analyzing the results of the base and strategy 1 on semi-cleaned data, no significant differences in F1-score were observed. Consequently, it prompted for further data cleaning, leading to improved outcomes. Notably, the best performance was achieved when the data was cleaned and the first three layers were freezed. This approach resulted in an F1-score of   0.72, 0.89, 0.73, 0.7, and 0.72 for the  disfluency types: word repetition, sound repetition, prolongation, interjection, and blocks, respectively. Moreover, the average weighted F1-score achieved was 0.81. We also explored the Whisper tiny model, which consists of four encoder layers. The outcomes indicated a significant degradation of F1-score. Consequently, we discontinue further experimentation with this version of Whisper.
\begin{figure}[h]
    \centering
     \includegraphics[width=\linewidth]{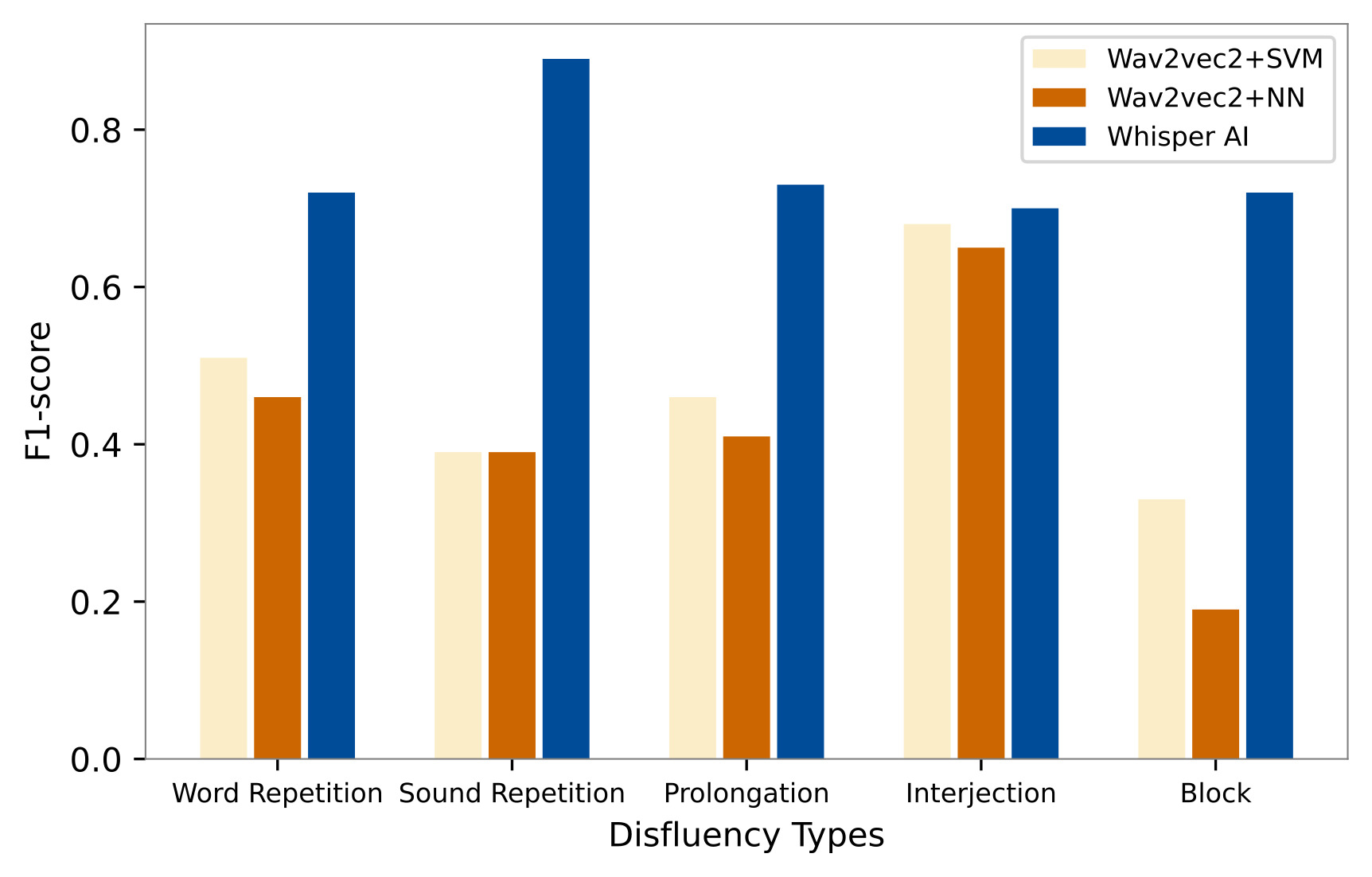}
    \caption{A Comparative Analysis of F1 Scores between the Whisper Model and Wav2vec2.0 + SVM/NN Models: A Demonstration of our Findings in Relation to a Prior Research Investigation \cite{yildirim2009automatic} }
    \label{fig:performance_img}
\end{figure}

Figure~\ref{fig:performance_img}, presents a comparison between our research findings and the model proposed by \cite{yildirim2009automatic}. Their methodology has leveraged Wav2vec2.0 in conjunction with two distinct classifiers i.e.  SVM and NN. Their results are reported on the test split of SEP-28k-E dataset. In contrast, through preprocessing and data cleaning steps, coupled with an optimized Whisper base model, we have attained state-of-the-art results during external testing. These outcomes firmly support the notion of \textit{Data quality over mere quantity}.

    Also, it is worth highlighting that even though Wav2vec2.0 and Whisper are two distinct models, their fine-tuning in this task occurs within the confines of a labeled dataset, and same loss functions i.e. cross entropy. In the context of this downstream task, the number of encoder layers were varying, along with their sequence lengths. However, their dissimilarity in feature extraction methodologies can be observed in Figure~\ref{fig:wav2vec2} and ~\ref{fig:whisper}. In Figure~\ref{fig:wav2vec2}, the Wav2vec2.0 model takes the raw audio which is standardized to zero mean and unit variance. It is then fed to seven block feature encoders each consisting of 1D convolutional neural networks  followed by layer normalization and Gaussian Error Linear Unit (GELU) activation function. In contrast, the feature extracted illustrated in Figure~\ref{fig:whisper}, Whisper converts the audio into log-Mel spectrogram representation first, and then subsequently subjecting it to two 1D convolutional layers. This difference in feature extractors may well account for Whisper’s superior performance relative to Wav2vec2.0. 
\begin{figure}[h]
    \centering
    \includegraphics[width=\linewidth]{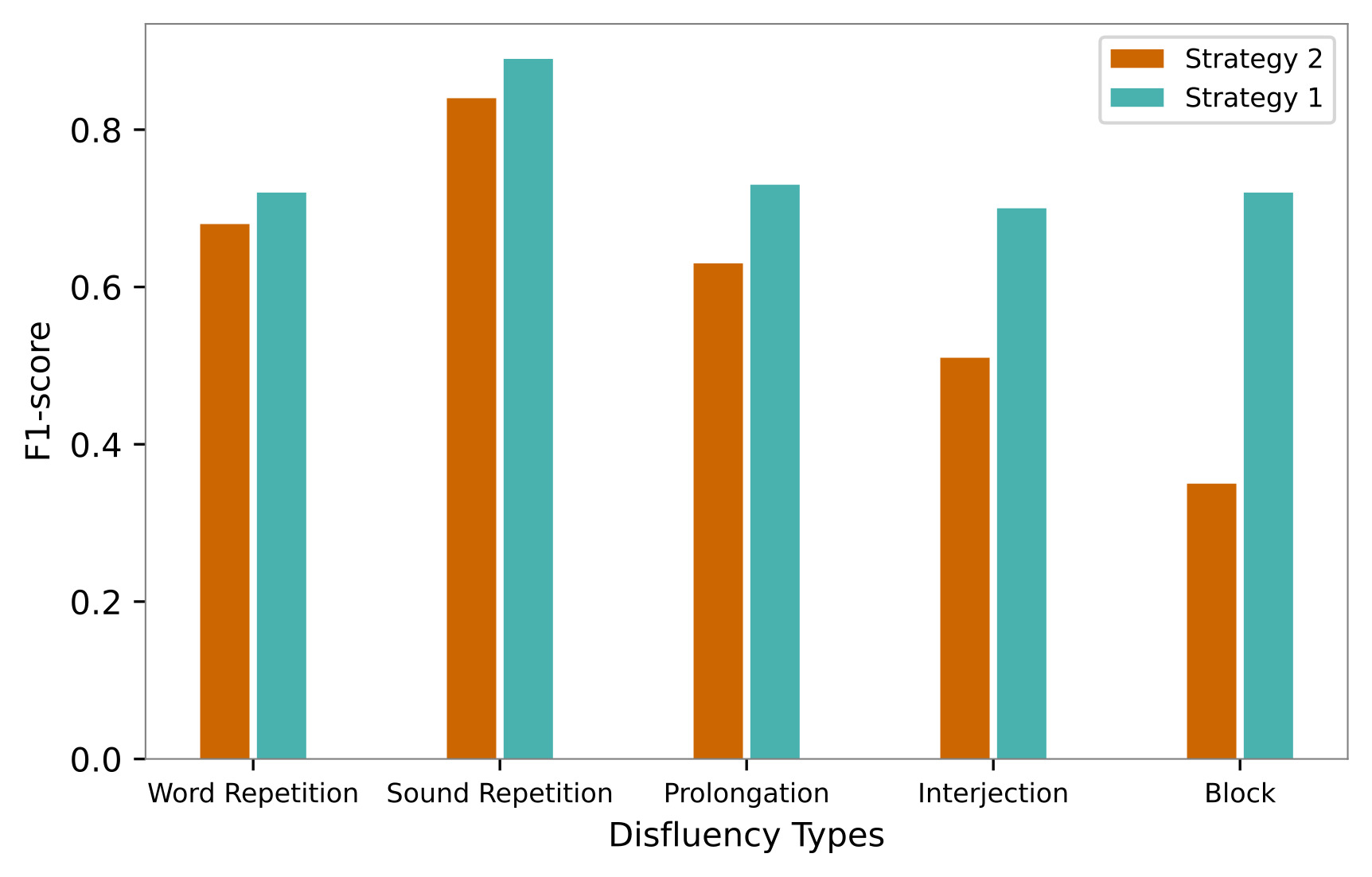}
    
    \caption{An Empirical Analysis: Evaluating the Efficacy of Strategies 1 and 2 through F1 Scores Across Disfluency Types}
    \label{fig:freeze_performance_img}
\end{figure}

A comparative analysis, as illustrated in Table~\ref{tab:my-result-table}, highlights the significance of freezing the layers. It is evident from the outcome that the base strategy results in equal or inferior performance compared to strategy 1. Additionally, a noticeable degradation in results can be observed by employing strategy 2, as demonstrated in Figure~\ref{fig:freeze_performance_img}. In Table~\ref{tab:my-result-table} it can be observed that the F1-score is reduced to 0.68, 0.84, 0.63, 0.51, 0.35 disfluency types: word repetition, sound repetition, prolongation, interjection, and blocks, correspondingly. This outcome indicates that the learned features in the deeper layers play a pivotal role in the better classification of disfluencies. A key advantage of freezing layers is the reduction in the number of trainable parameters as shown in Table~\ref{tab:model-parameters}. These parameters were reduced by 46\% in comparison with Whisper based model where all the layers were kept unfrozen. This approach has significantly improved the efficiency of the training process, making it more applicable in real-time.
\begin{table}[h]
\centering
\caption{Comparison of Trainable Parameters}
\label{tab:model-parameters}
\begin{tabular}{@{}l c@{}}
\toprule
\textbf{Model} & \textbf{Trainable Parameters (Millions)} \\
\midrule
Wav2vec2-base & 94.57 \\
Whisper-base & 20.72 \\
\textbf{Whisper-base (Strategy 1)} & \textbf{11.27} \\
\bottomrule
\end{tabular}
\end{table}

In contrast to the prior studies, where Wav2vec2.0 with 12 encoder layers have been used to achieve the classification task. In this study we are utilizing the Whisper base model which consists of 6 encoder layers. Despite this reduction in complexity, Table~\ref{tab:my-result-table} demonstrates that the best-performing configuration was achieved with strategy 1 when using cleaned version of the data. Therefore, without compromising the performance of the model, a more resource-efficient model can be employed. In addition, the difference in training runtime of Wav2vec2.0 and Whisper is substantial (Table~\ref{tab:training-runtimes}), making the latter a better choice for the task at hand. 
\begin{table}[h]
\centering
\caption{Comparison of Training Runtime of Speech Models}
\label{tab:training-runtimes}
\begin{tabular}{@{}lc@{}}
\toprule
\textbf{Model} & \textbf{Training Runtime (Seconds)} \\
\midrule
Wav2vec2.0 & 5995.19 \\
\textbf{Whisper} & \textbf{1389.07} \\
\bottomrule
\end{tabular}
\end{table}

Hence, we can conclude that Whisper is a more generalized model as in this study it has been fine-tuned for multi-class classification tasks for disfluency types. In contrast with the previous studies, the problem was considered as a binary classification task i.e. model was trained on distinct disfluencies. Secondly, with better outcome of freezed layers, Whisper is a more resource-efficient choice for this task as it reduces the computational overhead. Lastly, our results also identify the contributions of different layers with deeper layers contributing more towards better F1-score which open doors for further investigations. 
Nevertheless, there is still a need for a model which is generalized well enough to cater multilingual use case scenarios.

\section{Conclusion}
\label{sec:conclusion-future}
This research study set out to assess the potential of Whisper in the context of disfluency type classification of stuttered speech. Our primary objective centered on evaluating the generalizability of the transformer-based model, and resource efficiency. Given the intricate nature of stuttered speech identification, we have  delved into transformer-based models as they have yielded better results in the previous research studies. To the best of our knowledge Whisper has not been previously applied to this specific task. Its superior performance in other scenarios i.e. speech recognition, detection of deep fake, classification of vocal intensity prompted our investigation. In this use case, it has also achieved average weighted F1-score of 0.81, and an F1-score exceeding 0.7 for all disfluency types despite being evaluated on external dataset known as FluencyBank. This noteworthy performance was attained by fine-tuning Whisper on our enhanced version of SEP-28k dataset and selectively freezing the initial three encoder layers out of six. It  effectively reduced the number of  trainable parameters, making the model more optimized. As we look ahead, we aspire to extend this model’s evaluation to other spoken languages, broadening its applicability. Also, due to limited available data, the model could not be fine-tuned for scenarios in which disfluency types can co-occur. We aim to introduce an optimized model which is capable of handling multi-stuttered and multilingual use cases. This study marks a stepping stone, redirecting attention towards more efficient solutions which can be utilized in real-time application, promising innovation within the domain.

\section*{Acknowledgement}
\label{sec:acknowlegement}
The authors express their sincere appreciation to World Technology Partners, USA, for their grant to conduct this research. The authors also acknowledge the invaluable support and resources provided by the CPInS Lab at SEECS-NUST and the Artificial Intelligence and Data Analytics (AIDA) Lab at Prince Sultan University, which has been instrumental in facilitating the research and publication of this work.

\bibliographystyle{unsrtnat}
\bibliography{references}  






\end{document}